# Spatially, Temporally and Polarization-Resolved Photoluminescence Exploration of Excitons in Crystalline Phthalocyanine Thin Films

Naveen Rawat,[†] Zhenwen Pan,[†,&] Lane W. Manning,[†] Cody J. Lamarche,[†] Ishviene Cour,[†,$] Randall L. Headrick,[†] Rory Waterman,[‡] Arthur R. Woll,[§] and Madalina I. Furis[†,*]

[†]Department of Physics and the Material Science Program, University of Vermont, Burlington, Vermont 05405, United States

[‡]Department of Chemistry, University of Vermont, Burlington, Vermont 05405, United States

[§]Cornell High Energy Synchrotron Source, Cornell University, Ithaca, New York 14853, USA

AUTHOR INFORMATION

**Corresponding Author**

* Email: Madalina.Furis@uvm.edu

**Present Addresses**

[&]CGG, Houston, TX

[$]Intel Corporation, Portland, OR

**ABSTRACT**: The lack of long range order in organic semiconductor thin films prevents the unveiling of the complete nature of excitons in optical experiments, because the diffraction limited beam diameters in the bandgap region far exceed typical crystalline grain sizes. Here we present spatially-, temporally- and polarization-resolved dual photoluminescence/linear dichroism microscopy experiments that investigate exciton states within a single crystalline grain in solution-processed phthalocyanine thin films. These experiments reveal the existence of a delocalized singlet exciton, polarized along the high mobility axis in this quasi-1D electronic system. The strong delocalized orbitals overlap controlled by the molecular stacking along the high mobility axis is responsible for breaking the radiative recombination selection rules. Using our linear dichroism scanning microscopy setup we further established a rotation of molecules (i.e. a structural phase transition) that occurs above 100 K prevents the observation of this exciton at room temperature.

**TOC GRAPHICS**

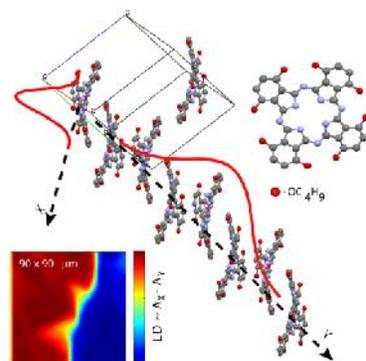

In recent years, ultrafast spectroscopy techniques have emerged as indispensable tools for studying the optical and electronic properties of organic thin films.[1-3] Further development in the field of Organic Light Emitting Devices (OLEDs), improvement in organic solar cell efficiencies and advancement in synthetic chemistry each demand a better understanding of excitonic states in organic thin films, especially in the presence of stronger intermolecular interactions.[4-6] The unique properties of organic semiconductors combined with recently-developed laterally directed solution processing methods make them a viable alternative to their inorganic counterparts.[7-10] Fundamental studies on excitonic properties of single crystals of small molecule organic thin films are still rare due to the polycrystalline nature of samples inherent to standard fabrication methods that result in a high level of disorder.[7, 11-14]

Time Resolved Photoluminescence (TRPL) experiments directly probe the excitonic radiative recombination and relaxation dynamics. These measurements give vital information about excited states, exciton binding energy and exciton recombination rates since they can selectively probe electronic states at or around the band gap. TRPL experiments are also highly sensitive to any disorder present in the system and, therefore, often limited in their usefulness because the typical grain sizes are orders of magnitude smaller than diffraction -limited focused laser beams diameters.[15, 16] Phthalocyanines (Pc's) are promising organic semiconductor candidates that can overcome these limitations provided large enough grain sizes can be obtained in a thin film sample. These small molecules were shown to exhibit high mobilities and long range interactions.[17-21]

In this letter, we present results of spatially-, temporally-, and polarization-resolved photoluminescence spectroscopy experiments that examine radiative exciton recombination within a single crystalline grain, in thin films of two soluble metal-free substituted

phthalocyanines: 1,4,8,11,15,18,22,25-octabutoxy-29$H$,31$H$-Phthalocyanine (H$_2$-OBPc) and 2,3,9,10,16,17,23,24-Octakis(octyloxy)-29$H$,31$H$-Phthalocyanine (H$_2$Pc-OC$_8$). These films were fabricated on c-plane sapphire substrates using an in-house developed, pen-writing technique,[10, 22] described in the Supporting Information. The films are a good example of a quasi-1D electronic system where the delocalized -orbital overlap is highly directional, generating a specific high carrier mobility axis in the crystal. The highly anisotropic behavior of the ordered thin films and their impact on carrier mobility and exciton diffusion lengths has been recently established.[22-26] Moreover, we developed a Linear Dichroism (LD) laser scanning microscopy experiment which enables us to investigate molecular ordering and its effects on excitonic states.[27]

Figure 1a shows the polarized optical microscope images of our solution processed H$_2$-OBPc thin films used in this study. Similar images (Figure S1) were also acquired from films of another soluble derivative, H$_2$Pc-OC$_8$. In both cases, the carbon chains attached to the Pc ring improve solubility and influence the crystalline structure where molecules stack in a quasi-1D chain along a particular crystalline direction, as depicted in Figure 1b. Selection rules dictate that light polarized along this stacking direction is not absorbed, hence the term "dark axis" often employed for the stacking axis. Hence, the strong contrast seen in Figure 1a originates in the different orientation of stacking axis relative to the polarizer/analyzer axes of the microscope. As the film is rotated, this relative orientation changes and different grains become bright. It is important to note the grains are mm-sized, which implies the long range order is preserved over the entire area of a laser beam (less than 5 μm in diameter) used to probe these films. Figure 1b depicts the known powder crystal structure for H$_2$-OBPc with the stacking-axis denoted by the dashed arrow.[28] The crystal symmetry is triclinic with two molecules per unit cell denoted by p$_1$ and $p_2$. The lattice constants a, b, and c, as well as the nearest neighbor distances along the stacking axis, $d_1$ and $d_2$, are also marked in the same figure. It is noteworthy that the introduction of the side chains leads to a significantly different crystalline structure and packing than the better known insoluble parent Pc molecule, where the stacking axis is the c-axis of the unit cell.[29] Specifically, the side chains lead to very different stacking angles (i.e. the angle made by the molecule plane with the stacking axis). However, regardless of the side chain, the nearest neighbor distances along the stacking axis and the 1D character are very similar.

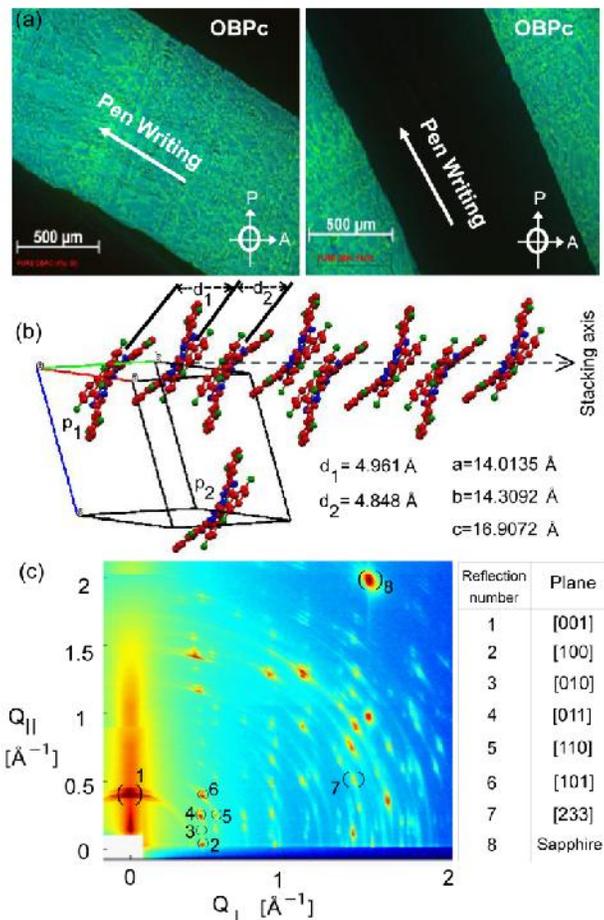

**Figure 1.** (a) Polarized microscope images of an $H_2$-OBPc thin film showing long range ordering. The contrast originates from the different grain orientations with respect to the microscope polarization axis. (1 wt. % in toluene at 0.01mm/sec; Scale bar is 500 μm.) (b) $H_2$-OBPc crystal structure[28] (the octabutoxy group was omitted for clarity). The dashed arrow points along the molecular stacking axis [110] (c) 2D reciprocal space maps recorded in the GIXRD experiment on an $H_2$-OBPc thin film. The table lists our reflections assignments that match previously reported values.[28]

The crystallinity of the film was confirmed through a grazing incidence X-ray diffraction (GIXRD) experiment where the sample was rotated around the axis normal to the substrate to capture as many different reflections as possible and compare them to the known powder XRD set of reflections. Figure 1c depicts GIXRD data for a $H_2$-OBPc thin film. Well-defined, sharp reflections were indexed according to known XRD data (see table in Figure1c and table SII). Our films exhibit very good crystallinity and long range order with the stacking axis parallel to the substrate (c-axis pointing out of the plane of the substrate).

Figure 2a shows a schematic of the LD/absorption/TRPL measurement geometry along with the typical molecular ordering of the thin film on the substrate. **K$_{in}$** refers to the direction of incident light, **E$_x$** and **E$_y$** relate to the electric field vectors of the incident light polarized along the **x**- and **y**- lab axes, whereas **K$_{PL}$** refers to the luminescence collected in backscattering geometry. Absorption and LD are simultaneously recorded in the transmission geometry. The absorbance spectrum measured in solution (Figure 2b) exhibits a splitting of the peak (740 nm and 758 nm) associated with the HOMO-LUMO bandgap (Q-band) transition[30] that originates in the D$_{2h}$ symmetry of these molecules. The thin film absorbance spectrum is significantly harder to assign since the crystalline order promotes unidirectional long range interactions between molecules along the stacking axis, giving rise to new exciton states in the band gap manifold.[31-34] Furthermore, the broadening of the bands due to small Davydov splitting and excitonic coupling complicates the interpretation of the observed absorbance spectrum,[35-37] as indicated by the band gap in our $H_2$-OBPc thin film which is red-shifted to 870 nm. The appearance of additional electronic states is also evident in the LD spectrum that features three shoulders in the Q-band region. Linear dichroism, defined as the difference in absorbance ($A_X$ - $A_Y$) for light polarized parallel and perpendicular to the molecular plane,[38] is observed in the thin films because the selection rules dictate that light polarized along the stacking axis is not absorbed. In films such as the one presented in Figure 1a, the orientation of the stacking axis remains largely the same over the entire illuminated area, hence the large amount of observed LD. The same selection rules account for the large difference between the **x**- and **y**- polarized PL intensities, $I_X$ and $I_Y$, because exciton transition dipoles oriented along the stacking axis are nominally forbidden in the "oriented gas model".[39, 40] However, in crystalline thin films, we expect that delocalization and stronger long range interactions along the stacking axis will result in new selection rules stemming from the crystalline rather than molecular symmetry.

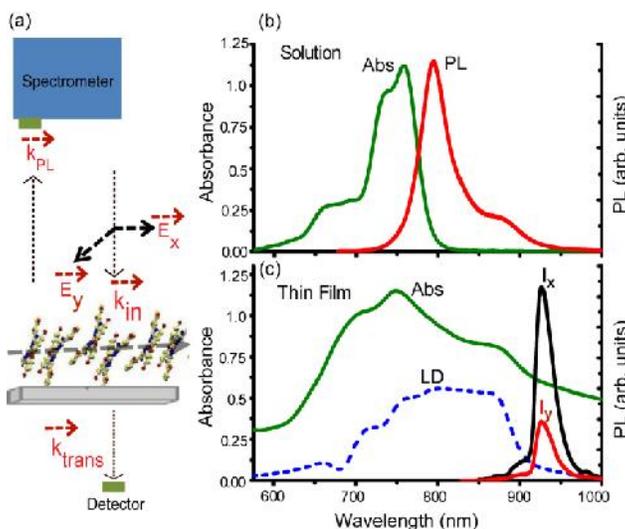

**Figure 2.** (a) Schematics of the PL/Absorption/LD experiment employed to study the correlation between excitons and edge-on molecular ordering of H$_2$-Pc thin films (b) Absorption and PL emission of H$_2$-OBPc in solution (c) Absorption and Linear Dichroism spectra from an H$_2$-OBPc crystalline thin film along with the **x**- and **y**-polarized components, **I$_x$** and **I$_y$**, of the PL spectrum from a single grain at 5 K.

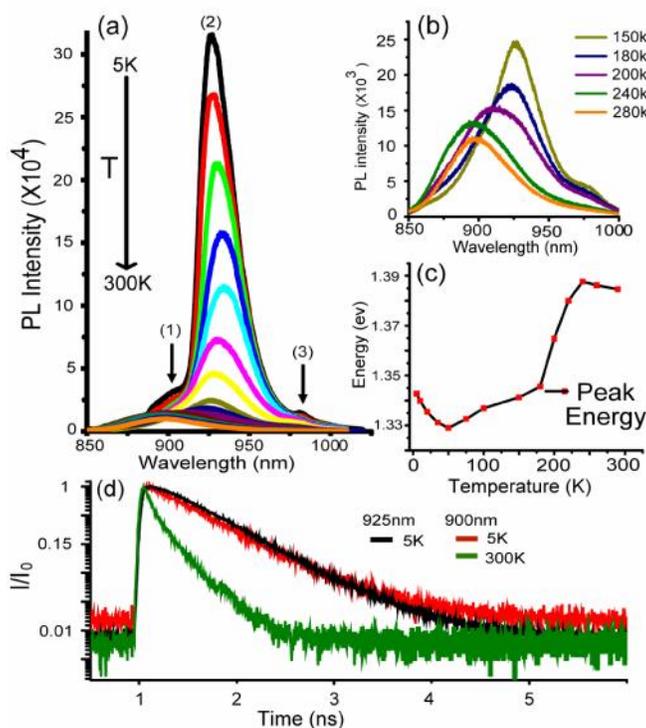

**Figure 3**. (a) Temperature evolution of the PL spectra of crystalline $H_2$-OBPc film (b) PL spectra recorded above 150 K clearly indicates the blue shift of luminescence at 180 K. (c) Temperature dependence of PL peak energy with temperature (d) PL decay measured at 5 K and 300 K for features (1) and (2).

The temperature dependence of exciton recombination in crystalline $H_2$-OBPc thin films, summarized in Figure 3, confirms this expectation. Figure 3a shows the evolution of the PL spectra with temperature from 5 K to 300 K. The spectrum first red shifts with increasing temperature and it is not until approximately 180 K that we observe a large blue shift (see Figure 3b,c). This evolution can be better understood by fitting the low temperature spectra with three Gaussian features, labeled (1), (2) and (3) in Figure 3. Details of our fitting procedure can be found in the Supporting Information. For temperatures larger than 180 K the spectrum is dominated by a single Gaussian feature, as illustrated in Figure S3. This evolution clearly indicates the existence of a thermally-activated switch that turns on the red-shifted excitonic state (feature 2) below 180 K. For T > 180 K, the PL emission is resonant to the absorption edge, and its intensity and polarization remain almost constant throughout the entire temperature range, identifying it as the optically-allowed singlet exciton recombination. In contrast, feature (2), that heavily dominates the spectrum at 5 K, exhibits a dramatic drop in intensity with increasing temperature before disappearing rather abruptly around 180 K. Feature (3) is the least intense in the spectrum and evolves with temperature identically to feature(2). For that reason we interpret it as a phonon/vibronic replica of feature (2).

A simple oriented gas model for the excitons in an 1D molecular chain[40] predicts there are two excitonic states at the bandgap: one polarized along the stacking axis and the other perpendicular to it. The former is optically-forbidden (dark) in this approximation while the latter is always allowed (bright). Their energies are highly dependent on the angle the molecular plane makes with the stacking axis, which also determines whether the ground state is dark or bright. This model assumes periodic boundary conditions and no interactions between molecules or orbital overlap along the stacking axis. We hypothesize that feature (2) represents the radiative recombination of this "dark" state, that is now optically allowed because new selection rules dictated by crystalline symmetry, interactions along the chain, and the long range order apply instead. The strong temperature dependence in Figure 3a is in accordance with this picture, since the luminescence behaves very much like bulk exciton recombination with a Full Width at Half Maximum (FWHM) increasing with temperature due to increasing presence of phonons.[41]

More recent theoretical studies of Spano and co-workers,[42-45] showed that strong exciton/vibronic coupling and intermolecular excitonic interactions can lead to significant exciton delocalization and the formation of delocalized polarons in small molecule thin films.

We propose that in the case of phthalocyanines this "bulk" exciton wavefunction is delocalized along the molecular stacking axis. Further evidence in support of this hypothesis is provided through low temperature time -resolved PL measurements of exciton lifetimes. At 5 K, the PL decay at 925 nm is characterized by a lifetime of approximately 1 ns, very similar to the singlet "bright'", exciton radiative recombination lifetime (900 nm) at the same temperature (Figure 3d). This two-exciton (one localized, one delocalized) picture changes dramatically around 180 K, where a significant blue shift in the PL spectrum (Figure 3c) and recombination from a single, temperature –independent exciton are observed.

We confirmed the existence of the two (localized and delocalized) exciton states in a similar system: an $H_2Pc-OC_8$ thin film deposited using the same pen-writing procedure. In this film, the peripherally attached eight-carbon chains should lead to weaker intermolecular interactions and less orbital overlap along the stacking axis.[46] The results of TRPL experiments for this film are summarized in Figure 4. At cryogenic temperatures (5 K), the spectrum was again easily fitted with three Gaussians, labeled (1) through (3). Feature (1), associated with the singlet exciton emission is entirely independent of temperature. Its energy, intensity, and linear polarization (40%) remain constant through the entire temperature range. This result confirms vibronic states are not involved in this transition.[47] In contrast, feature (2) dominates the spectrum at 5 K, and decreases dramatically in intensity with increasing temperature. Moreover, its polarization increases from approximately 20% at low temperatures to 40% at 300 K. Figures 4b,c show the energy and polarization of the feature (2) as a function of temperature, extracted from fittings of the **x**- and **y**- polarized components of the PL spectra. There is no significant change for either the energy or polarization until the temperature reaches 100 K. Above that temperature, both energy and polarization gradually increases with temperature while the feature loses intensity in favor of the singlet exciton emission. The PL decay time measured at 5 K for the 815 nm feature is 6 ns, twice as long as the singlet exciton (768 nm) recombination (see Figure 4d). The significantly longer lifetime, the reduction in linear polarization and the red shift with respect to the absorption edge indicate feature (2) is indeed associated with an optically forbidden transition that becomes allowed at low temperatures.

The interplay between dark and bright exciton recombination has been reported previously in a variety of systems, albeit with different origins. For example, in colloidal nanoparticles, one typically observes luminescence from a "dark" ground state at low temperatures that evolves into a blue shifted luminescence at higher temperatures, because of mixing with energetically close, bright excitons.[48, 49] Our observation could be interpreted in a similar way, however, the abruptness of the temperature switching mechanism is not consistent with bright-dark exciton mixing; it is reminiscent of a phase transition.

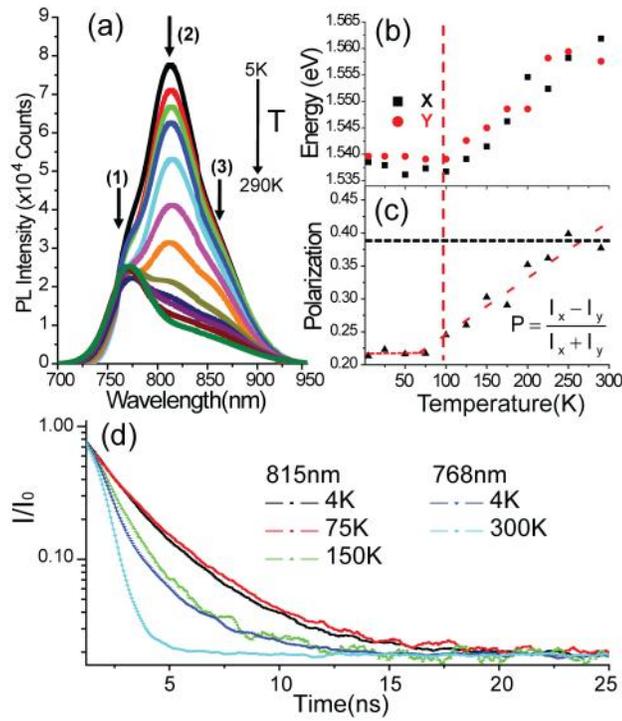

**Figure 4**. (a) Temperature evolution of the x-polarized component of the PL spectrum in crystalline $H_2Pc-OC_8$ film (b) Temperature dependence of the energy corresponding to **x**- and **y**-polarized components of feature (2) (c) Evolution of the linear polarization with temperature) (d) Temporal decay of features (1) and (2) recorded at various temperatures.

To establish the origin of the temperature-induced switching between localized and delocalized excitons, we employed a dual Linear Dichroism (LD)/PL laser scanning microscopy experiment, designed for grain boundary imaging to monitor the evolution with temperature of the $H_2$-OBPc molecular stacking in the thin film.[27] The LD imaging was performed on two adjacent grains, as shown in Figure 5, by raster scanning a focused laser beam across the sample surface and recording the LD signal pixel by pixel. For details of the experiment, please refer to the Supporting Information and reference 27. As discussed earlier, LD is defined as the differential absorption of linearly polarized light by an ordered sample and can be expressed as a function of molecular stacking[50] as follows:

$$LD = (Cos^2 \theta)(Sin^2 \varphi - Cos^2 \varphi) \quad (1)$$

where angle $\theta$ is the polar angle representing the tilt of the molecular plane normal with respect to the normal direction to the substrate (z-axis). The angle $\varphi$ is the azimuthal angle between the molecular stacking axis and the polarizer axis. The images show contrast across the grain boundary with LD values either positive or negative depending on the angle $\varphi$ for each of the grains. The data shows there is little change in LD signal until we reach the range of the 180 K – 200 K where LD starts decreasing dramatically. It is important to note that while the overall LD levels drop significantly, LD contrast remains essentially unchanged. That means the $cos^2 \theta$ term in equation (1) is responsible for this drop, suggesting the molecular tilt changes with temperature. A rough calculation following eq. (1) suggests the angle $\theta$ increases from approx. 25° at 5 K to about 65° at room temperature, while $\varphi$ remains constant. We therefore conclude the switching of excitonic states observe in Figure 3 is due to a change in molecular plane orientation relative to the stacking axis that, in turn, is responsible for a change in -orbital overlap.

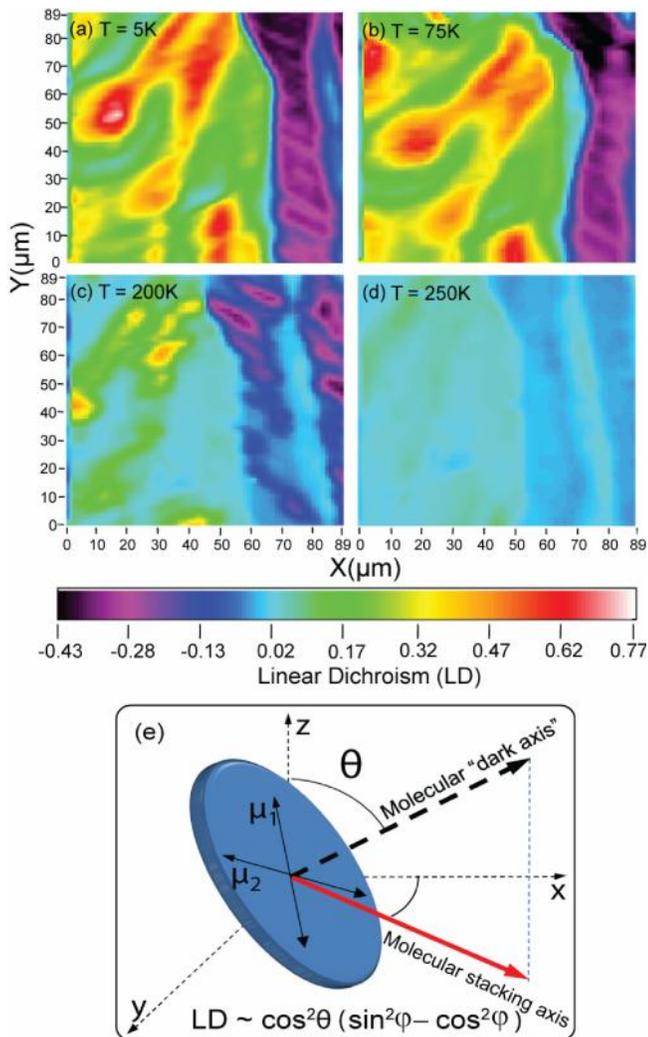

**Figure 5**. (a)-(d) LD scanning microscopy images of a 90 x 90 μm area of an $H_2$-OBPc thin film recorded at 5 K, 75 K, 200 K and 250 K respectively. The area chosen for imaging is traversed from top to bottom by a single grain boundary and the LD contrast originates from the different orientations of stacking axes in adjacent grains different temperatures (b) Stacking axis orientation and its relation to the LD signal.

In the framework of the simple 1D oriented gas model[40] of a molecular chain with two molecules per unit cell mentioned earlier, a straight forward calculation of the two bandgap excitons energies predicts the exciton ground state is an optically-allowed dipole oriented perpendicular to the stacking axis if the angle „ is larger than 54.2°. For „ values smaller than 54.2°, the ground state is an optically forbidden dipole oriented along the stacking axis. The measured energy difference between the two excitonic states (40 meV) is also in agreement with the range of values predicted by the same calculation (30 - 70 meV, depending on the angle „).

In conclusion, our combined LD/PL microscopy of metal-free Pc thin films revealed that the lowest excitonic singlet transition is linearly polarized, in agreement with predictions from a simple 1D molecular chain excitonic model. At low temperatures, the PL spectrum reveals the presence of a second, red-shifted exciton state that is delocalized along the stacking axis of the thin films. A temperature-induced change of molecular tilt with respect to the stacking axis acts like an on-off switch for the delocalized exciton state. From a broader perspective, the presence of a delocalized extended excitonic state is important for many applications such as photovoltaics and merits further exploration. The macroscopic grain sizes present in solution processed films and the added chemical design flexibility of the substituted molecules enabled us to show this exciton state could be potentially tailored through control of molecular packing. The simultaneous observation of two different exciton states at low temperatures also opens the opportunity for a direct comparison of their diffusion lengths, an essential parameter for photovoltaic applications.

While phthalocyanines represent a good test system for such investigations because they are resistant to oxidation and retain a high yield of excitonic recombination in crystalline films, the studies reported here open an avenue for the exploration of electronic correlations that span several lattice sites in other crystalline small molecule semiconductors. For example, there has been tremendous success in fabricating single-crystal thin film OFETs structures from soluble substituted pentacene[7, 51] and our group has very recently deposited films of tetra-phenyl porphyrin (H$_2$TPP), octabutoxy naphthalocyanines, and a variety of soluble substituted metal phthalocyanines with grain sizes similar to Figure 1. The possibility of observing and controlling the intermolecular phenomena in ordered systems through molecular design, combined with rapid feedback optical spectroscopy techniques, can be an effective solution for tailoring optical and electronic properties of these organic semiconductors.

EXPERIMENTAL SECTION

The starting materials and solvents used were purchased from Sigma Aldrich. 1,4,8,11,15,18,22,25-octabutoxy-29*H*,31*H*-Phthalocyanine (H$_2$-OBPc) and 2,3,9,10,16,17,23,24-Octakis(octyloxy)-29*H*,31*H*-Phthalocyanine (H$_2$Pc-OC$_8$) powders were purified by column chromatography followed by recrystallization. Both of these Pc derivatives show excellent solubility in common organic solvents after purification. Crystalline thin films were deposited on c-plane cut sapphire substrates using a hollow capillary pen-writing technique at room temperature.[10, 22] This technique produces macroscopic grain sizes under optimum deposition conditions. GIXRD was performed at the CHESS Synchrotron facility at Cornell University to confirm the crystalline structure of our films. The excitation source used for TRPL experiments is an ultrafast Coherent Mira 900 pulsed laser used in conjunction with a frequency doubling crystal setup that delivers 2ps pulses at a repetition rate of 76 MHz and a photon wavelength of 400 nm. For longer radiative lifetimes, we employed a narrowband 405 nm Coherent CUBE diode laser electronically triggered to output 7 ns long pulses at a repetition rate of 8 MHz. The sample temperature was varied continuously from 5 K to 300 K using a continuous-flow Oxford Microstat He cryostat. A Picoharp time-correlated single photon counting (TCSPC) system from PicoQuant coupled to a Princeton Instruments Acton 0.5 m spectrometer was used to measure continuous wave PL spectra and spectrally-resolved PL decay. For further details on sample processing and experimental methods, please refer to the Supporting Information

ASSOCIATED CONTENT

**Supporting Information**

Experimental details, GIXRD calculations and spectral fitting procedures. This material is available free of charge via the Internet at http://pubs.acs.org.

AUTHOR INFORMATION

**Corresponding Author**

*Email: Madalina.Furis@uvm.edu

**Notes**

The authors declare no competing financial interests.

ACKNOWLEDGMENT

We would like to thank Dr. Anthony Wetherby for his assistance with purification of the materials. This work was supported by the National Science Foundation, Division of Materials Research MRI, CAREER and EPM program awards: DMR- 0722451, DMR- 0821268, DMR-1307017 and DMR-1056589.